# Complete steric exclusion of ions and proton transport in two-dimensional water

K. Gopinadhan,[1,2,3] S. Hu,[2,4] A. Esfandiar,[5,1] M. Lozada-Hidalgo,[1] F. C. Wang,[1,6] Q. Yang,[2,7] A. V. Tyurnina,[2,8] A. Keerthi,[1,2] B. Radha,[1,2] A. K. Geim[1,2]

[1]School of Physics & Astronomy, University of Manchester, Manchester M13 9PL, UK
[2]National Graphene Institute, University of Manchester, Manchester M13 9PL, UK
[3]Department of Physics, Indian Institute of Technology Gandhinagar, Gujarat 382355, India
[4]College of Chemistry and Chemical Engineering, Collaborative Innovation Center of Chemistry for Energy Materials, Xiamen University, Xiamen, 361005, China
[5]Department of Physics, Sharif University of Technology, P.O. Box 11155-9161, Tehran, Iran
[6]Chinese Academy of Sciences Key Laboratory of Mechanical Behavior and Design of Materials, Department of Modern Mechanics, University of Science and Technology of China, Hefei, Anhui 230027, China
[7]Key Laboratory of Advanced Technologies of Materials, School of Materials Science and Engineering, Southwest Jiaotong University, Chengdu, Sichuan 610031, China
[8]Skolkovo Institute of Science and Technology, Nobel St. 3, 143026 Moscow, Russia

*Biological membranes allow permeation of water molecules but can reject even smallest ions. Behind these exquisite separation properties are protein channels with angstrom-scale constrictions (e.g., aquaporins). Despite recent progress in creating nanoscale pores and capillaries, they still remain distinctly larger than protein channels. We report capillaries made by effectively extracting one atomic plane from bulk crystals, which leaves a two-dimensional slit of a few Å in height. Water moves through these capillaries with little resistance whereas no permeation could be detected even for such small ions as $Na^+$ and $Cl^-$. Only protons can diffuse through monolayer water inside the capillaries. The observations improve our understanding of molecular transport at the atomic scale and suggest further ways to replicate the impressive machinery of living cells.*

It has long been an aspirational goal to create artificial structures and devices with separation properties similar to those of biological membranes[1-5]. The latter utilize a number of separation mechanisms but it is believed that angstrom-scale constrictions within protein channels[6,7] play a key role in steric (size) exclusion of ions with the smallest hydration diameters $D_H \approx 7$ Å, typically present in biofluids and seawater[8,9]. Such constrictions are particularly difficult to replicate artificially because of the lack of fabrication tools capable to operate with such precision and, also, because the surface roughness of materials is typically much larger than the required angstrom scale[1]. Nonetheless, several artificial systems with nanometer and sub-nanometer dimensions were recently demonstrated, including narrow carbon and boron-nitride nanotubes[5,10,11], graphene oxide laminates[12,13] and atomic-scale pores in graphene and $MoS_2$ monolayers[3,4,14]. The resulting devices exhibited high selectivity with respect to certain groups of ions (for example, they blocked large ions but allowed small ones[12,13] or rejected anions but allowed cations and vice versa[2,3,5]). Most recently, van der Waals assembly of two-dimensional (2D) crystals[15] was used to make slit-like channels of several Å in height[16,17]. They were atomically smooth and chemically inert and exhibited little ($\leq 10^{-4}$ C cm$^{-2}$) surface charge[17]. The channels allowed fast water permeation[16] and blocked large ions with a complete cutoff for diameters larger than ~10 Å (ref. [17]). Small ions (for example, those in seawater with $D_H$ of ~7 Å) still permeated through those channels with little hindrance, showing that an angstrom-scale confinement comparable to that in aquaporins[6,7] is essential for steric exclusion of small-diameter ions. In this report, we describe 2D channels with the height $h$ of about 3.4 Å (ref. [18]), which are twice smaller than any hydrated ion (smallest ions are $K^+$ and $Cl^-$ with $D_H \approx 6.6$ Å)[8,19] but sufficiently large to allow water inside (effective size of water molecules is ~2.8 Å). The achieved confinement matches the size of protein constrictions in biological channels, a critical factor to mimic their selectivity[6,7].

Our devices were fabricated using the van der Waals assembly described previously[16,17]. In brief, two thin (~100 nm) atomically flat monocrystals were obtained by cleaving bulk graphite or hexagonal boron nitride (hBN) and placed on top of each other, with stripes of graphene serving as spacers between the two crystals (Fig. 1A). The resulting trilayer assembly can be viewed as a pair of edge dislocations with a 2D empty space connecting them. This space can accommodate only one atomic layer of water (Fig. 1B). The 2D channels were designed[16,17] to have the width $w \approx 130$ nm and the length $L$ of several µm. We normally used many channels in parallel ($\geq 100$) to increase sensitivity of measurements. The resulting structures were assembled on top of a silicon nitride membrane that separated two isolated containers so that 2D channels provided the only molecular pathway between the containers (Fig. 1A). For further details, see Supplementary Information. The principal difference with the capillaries reported earlier[16,17] is the use of monolayer graphene as spacers. Previously, water permeation could be discerned only for capillaries with thicker spacers ($h \geq 6.7$ Å) and molecular dynamics simulations also suggested that smaller 2D cavities should collapse because of van der Waals attraction between the opposite walls[16]. This has turned out to be incorrect as revealed by our Raman measurements, which show that water can fill even monolayer capillaries (Supplementary Fig. 1). To detect minute molecular flows through them, we needed to introduce the following improvements: 1) tenfold increase in the number $n$ of channels probed in parallel; 2) use thicker top crystals to avoid their sagging[16]; and 3) clip channels' mouths by plasma etching to remove their blockage by thin edges of the top crystal[18].

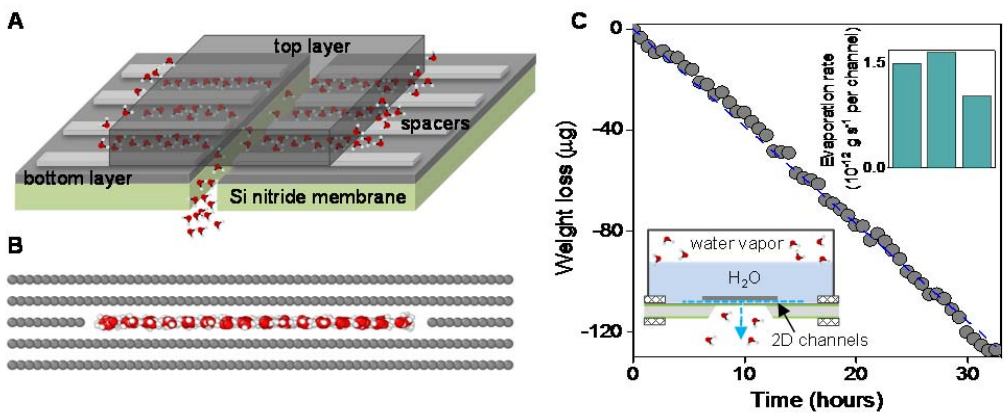

*Fig. 1. Transport of water through one-atom-high capillaries. (A) Schematic of our fluidic devices. Top and bottom layers are either graphite or hBN crystals. The spacers are monolayer graphene. The trilayer assembly is supported by a free standing membrane with a rectangular opening. (B) There is space only for a monolayer of water inside such 2D cavities as molecular dynamics simulations show[20,21]. (C) Weight loss due to water permeation through 1,500 channels and its subsequent evaporation; L = 2 µm. Bottom inset: Gravimetry setup. Top inset: Evaporation rates for three graphite/graphene/graphite devices with n = 1500. The shown rates are per 1 µm channel length. The detection limit of our gravimetry (~$10^{-13}$ g s$^{-1}$) is comparable to that of helium mass spectrometers and much better than that achieved by mass spectrometry for other gases and vapors using similar devices[16].*

We first describe water permeation through our capillaries. The inset of Fig. 1C shows a schematic of the used gravimetry setup[16]. The device containing 2D channels sealed a miniature container filled with water, and the weight loss was recorded as a function of time. For reference, we studied devices made in exactly the same way but without etching stripes within the graphene layer used for spacers[16]. No evaporation could be detected for the reference devices, even after many days of measurements. In contrast, the monolayer devices exhibited a clear weight loss evolving linearly in time (Fig. 1C). This translates into a water permeation rate of ~$1.5\times10^{-12}$ g·s$^{-1}$ per µm channel length per channel. Three such devices were studied, all showing similar rates (inset of Fig. 1C). This value is ~10 times smaller than the rates reported previously for devices with bilayer spacers ($h \approx 6.7$ Å) and is well below the detection limit achieved in ref. 16. The difference in water evaporation rates through mono- and bi- layer capillaries is larger than 4 times, which is expected from the $h^2$ dependence for the 2D channels[16,18]. The additional flow reduction (by a factor of 2) is not surprising because water becomes increasingly more viscous under strong confinement[16,22,23].

Having established that water can flow through our 2D channels, we investigated their permeability with respect to ions. In these experiments, the devices separated two reservoirs filled with chloride solutions (inset of Fig. 2A). The electrical conductance $G$ between the reservoirs was measured using chlorinated Ag/AgCl electrodes or calibrated reference electrodes[17] (Supplementary Fig. 2). First, we tested ion conductance using chloride solutions in 1 M concentrations. Within our detection limit of ~50 pS given by parasitic leakage currents of ~5 pA, no conductance could be detected for any salt (some of them are listed in Fig. 2A). This is in stark contrast to the behavior observed for capillaries with thicker spacers including bilayers[17]. The latter ($h \approx$ 6.7 Å) exhibited $G$ of the order of $10^{-8}$ S, that is, 3 orders of magnitude above our detection limit, in agreement with the conductance expected from the known conductivity $\sigma$ of the 1 M chloride solutions for the given length $L$ and cross-sectional area $n \times h \times w$ of the bilayer devices. The latter measurements were reported previously[17] but, for consistency, repeated in this work. Despite no discernable ion transport through capillaries with $h \approx$ 3.4 Å, they unexpectedly showed a substantial conductance if we used HCl rather than salt solutions (Fig. 2A, Supplementary Fig. 2). The measured $\sigma$ for the acid varied approximately linearly with its concentration $C$ as expected (Fig. 2B) but the absolute values were 10-30 times smaller than for bulk HCl (black curve in Fig. 2B). This is again in contrast to the behavior found for bilayer devices ($h \approx$ 6.7 Å), which exhibited the conductance that agreed well with that of bulk HCl in the given geometry (Fig. 2B).

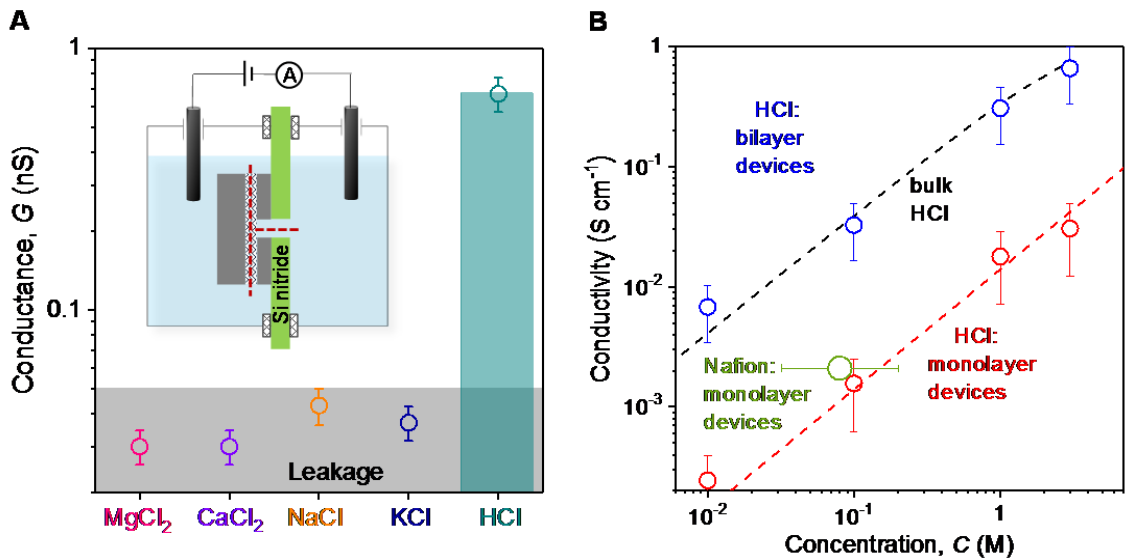

**Fig. 2. Steric exclusion from one-atom-high capillaries.** *(A) Conductance observed for one of our monolayer devices (n = 100; L = 10 µm; hBN/graphene/hBN) using the specified 1M solutions (symbols). The values of G were extracted from the slopes of I-V curves over a voltage interval of ±30 mV. Error bars show standard deviations (SD) in determining G from the curves. The grey area indicates our typical detection limit given by leakage currents, as found using both reference devices and blank Si nitride membranes. Several other common salts[17] were tested and exhibited no discernable conductance. In total, seven devices with monolayer spacers were studied showing the same behavior. For the case of HCl, G was well above our detection limit, although it varied between different devices from 0.1 nS to 0.64 nS (Supplementary fig. 2B) yielding on average G = 0.32 ± 0.2 nS. Inset: Schematic of our setup. (B) $\sigma$ calculated from the measured G for different HCl concentrations for mono- and bi- layer capillaries (red and blue symbols, respectively; red is the same device as in A). Red curve is the linear fit; black the literature values for bulk HCl. The green symbol shows $\sigma$ for monolayer devices using Nafion (instead of HCl) as proton reservoirs (see below). The horizontal error bar indicates the uncertainty in determining the proton concentration in Nafion films[24].*

The fact that no ion currents could be detected for salt solutions indicates steric rejection of all hydrated ions from the one-atom-high channels. It was previously reported[17] that ion permeation remained practically unaffected if $D_H \leq h$ but was sharply suppressed (~ 15 times) if $D_H$ exceeded $h$ by only 50%. In our case, $h \approx 3.4$ Å is at least twice smaller than $D_H$ of the tested ions, and the extrapolation of the previously observed cutoff (quicker than exponential)[17] suggests that the monolayer capillaries should completely prohibit transport of any ion. This conclusion does not contradict the finite $G$ observed for HCl because of a high concentration of protons in the latter case. They diffuse in water not as ions dressed in large hydration shells but more like particles jumping between water molecules, according to the Grotthuss mechanism[25,26]. In short, protons do not have hydration shells with a finite $D_H$ and, therefore, are not excluded by the angstrom confinement.

To corroborate the suggested proton transport scenario, we used the same ion measurement setup but different HCl concentrations in the two reservoirs. The concentration gradient drives both H$^+$ and Cl$^-$ in the same direction, from the high concentration ($C_h$) reservoir to the low concentration ($C_l$) one, and the sign of the electric current $I$ at zero applied voltage $V$ indicates whether majority carriers are H$^+$ or Cl$^-$ (refs. 8,17). Examples of $I$-$V$ characteristics observed for our monolayer devices are shown in Fig. 3A. One can see that the current at zero $V$ was always positive, indicating the dominant proton transport. Moreover, for the chosen gradient $\Delta = C_h/C_l = 3$, the $I$-$V$ curves intersect the current axis at the same $V_0 = 28 \pm 4$ mV regardless of the used $C_l$ and $C_h$. This voltage that compensates the currents caused by concentration gradients is described by[27]

$$V_0 = (t_H - t_{Cl})(kT/e) \ln(\Delta)$$

where $t_{H,Cl} \in [0,1]$ are the so-called transport numbers for H$^+$ and Cl$^-$, $k$ is the Boltzmann constant, $T$ is the temperature (295±3 K in our experiments), and $e$ is the elementary charge. For $\Delta = 3$ and $V_0 \approx 28$ mV, the formula yields $t_H - t_{Cl} \approx 1$, which can only be satisfied if $t_H \approx 1$ and $t_{Cl} \approx 0$. This means that only protons diffuse through our monolayer devices whereas Cl$^-$ ions are rejected, in agreement with the conclusions reached from the experiments of Fig. 2.

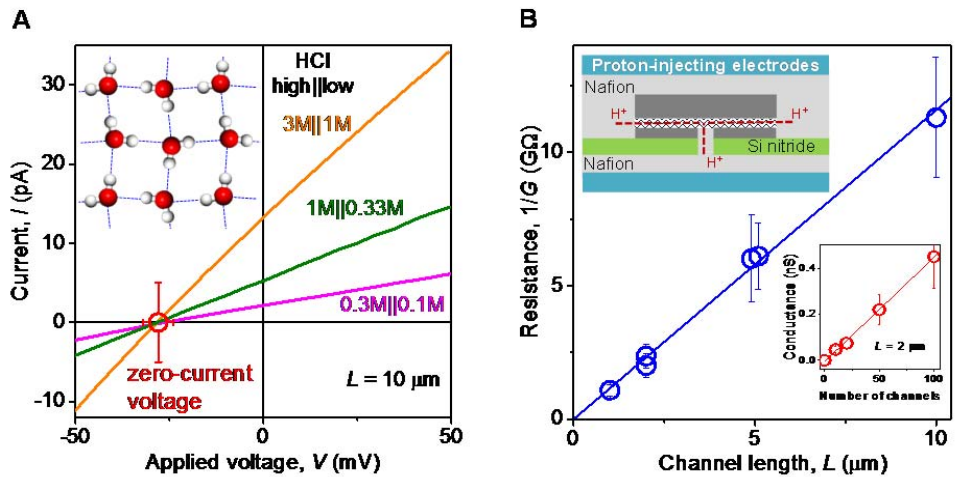

**Fig. 3. Proton transport through monolayer water inside 2D channels.** *(A) I-V curves for different concentrations of HCl in the two reservoirs connected by monolayer capillaries. Same $\Delta$ for all the curves (color coded). The red symbol marks voltages at zero I. The vertical bar indicates our maximum leakage currents; horizontal bar is the statistical error in determining $V_0$ using measurements for several monolayer devices. Inset: Square-like ice is one of the possible ices expected inside the 2D channels at room temperature, as found from molecular dynamics analyses[18,28,29]. (B) Proton transport using Nafion as proton reservoirs. Top inset: Schematic of our devices. Blue symbols: Their resistance as a function of length (hBN/graphene/hBN; n = 100). Bottom inset: G(n) for several Nafion devices with L = 2 µm. Red and blue lines: Best linear fits. Error bars are SD, as in Fig. 1.*

Proton transport can be probed more directly using Nafion instead of HCl (ref. 24). Nafion is a polymer in which protons are the only charge carriers and the counterions ($SO_3^-$) are fixed within the polymer matrix[30]. Using the approach described in ref. 24, we coated our devices with Nafion from both sides, and hydrogenated Pt films were used as proton-injecting electrodes converting electric current into a proton flow (top inset of Fig. 3B). For electrical measurements, these devices were placed in a hydrogen atmosphere at 100% relative humidity to ensure high proton conductivity of Nafion[24]. In comparison with our ion transport measurements, the Nafion devices were much more robust, sustained higher $I$ (ref. 18), exhibited smaller leakage currents (~ 1 pA) and showed little variations between similar devices. Because polymer molecules are much larger than $h$, they cannot enter 2D channels and, therefore, Nafion served in our experiments only as a reservoir providing protons for monolayer water inside the channels. Fig. 3B shows examples of the electrical resistance $R = 1/G$ exhibited by our devices. It varied linearly with $L$ and $1/n$, as expected. Reference devices with no spacers showed negligible conductance (~ $10^{-11}$ S; error bars in Fig. 3B). One can see that $R$ accurately extrapolates to zero in the limit of zero $L$ (blue curve in Fig. 3B), which yields that protons experience little barrier for entry into the capillaries. The experiments yield $\sigma \approx 2.1 \pm 0.2$ mS cm$^{-1}$ whereas the proton concentration in Nafion and, hence, inside the 2D water can be estimated from conductivity of our Nafion coating. It was found to be ~70 mM but with a relatively large uncertainty given by details of Nafion preparation[24]. These values (green symbol in Fig. 2B) agree well with $\sigma$ found from the HCl measurements, which further substantiates our conclusion of proton-only transport through one-atom-high channels.

It is instructive to estimate the diffusion constant $d_p$ of protons within monolayer water that fills the capillaries. It is given by $d_p = \sigma(kT/eFC_p) \approx 4\times10^{-6}$ cm$^2$ s$^{-1}$ where $C_p$ is the proton concentration in HCl or Nafion, and $F$ is the Faraday constant[18]. This value is more than an order of magnitude smaller than $d_p \approx 9\times10^{-5}$ cm$^2$ s$^{-1}$ for bulk water[9,26] which is also obvious from the difference between the red and black curves in Fig. 2B, if one recalls that the conductivity of bulk HCl is ~ 80% dominated by protons[8]. On the other hand, proton transport in 3D water is several times slower than that in 1D water inside sub-nm carbon nanotubes[5,11], which exhibited $d_p \approx 4\times10^{-4}$ cm$^2$ s$^{-1}$. The anomalously slow proton diffusion in 2D and nonmonotonic dependence of $d_p$ on dimensionality seem puzzling ($d_p$ in 2D is much smaller than in both 1D and 3D) but are consistent with different hydrogen bonding for the three cases[18]. For example, water molecules under 2D confinement may form either hexagonal or square-like ice at room temperature[20] (the latter is illustrated in Fig. 3A). The highly-ordered hydrogen bonding in 2D water should suppress rotation of water molecules with respect to 3D water, which is an essential but slowest step for proton transport, according to the Grotthuss mechanism[25,26]. This is in contrast to 1D water that exhibits a string-like hydrogen bonding, which was argued[5,11] to enhance proton transport (Supplementary Fig. 3).

To conclude, our one-atom-high voids allow one monolayer of water inside but exclude all ionic species. This is in contrast to the behavior of twice larger slits ($h \approx 6.7$ Å) which showed little effect on such small ions as Na$^+$ and Cl$^-$ ($D_H \approx 7$ Å). The extreme steric exclusion for $h \approx 3.4$ Å is attributed to the fact that it is no longer possible for any ion to squeeze inside the slits by deforming their hydration shells. Protons are different in that they diffuse not as hydrated H$^+$ ions but follow the Grotthuss mechanism. The slow diffusion observed for protons in 2D water suggests that it forms a proton ordered ice, in agreement with molecular dynamics simulations.

**References**

1. R. B. Schoch, J. Han, P. Renaud, Transport phenomena in nanofluidics. *Rev. Mod. Phys.* **80**, 839-883 (2008).
2. F. Fornasiero *et al.*, Ion exclusion by sub-2-nm carbon nanotube pores. *Proc. Natl. Acad. Sci. U.S.A.* **105**, 17250-17255 (2008).
3. S. C. O'Hern *et al.*, Selective ionic transport through tunable subnanometer pores in single-layer graphene membranes. *Nano Lett.* **14**, 1234-1241 (2014).

4. T. Jain *et al.*, Heterogeneous sub-continuum ionic transport in statistically isolated graphene nanopores. *Nat. Nanotechnol.* **10**, 1053-1057 (2015).
5. R. H. Tunuguntla *et al.*, Enhanced water permeability and tunable ion selectivity in subnanometer carbon nanotube porins. *Science* **357**, 792-796 (2017).
6. K. Murata *et al.*, Structural determinants of water permeation through aquaporin-1. *Nature* **407**, 599-605 (2000).
7. H. Chen *et al.*, Charge delocalization in proton channels, I: the Aquaporin channels and proton blockage. *Biophys. J.* **92**, 46-60 (2007).
8. B. Hille, *Ion channels of excitable membranes*. (Sinauer, Sunderland, MA, ed. 3, 2001).
9. C. A. Wraight, Chance and design—Proton transfer in water, channels and bioenergetic proteins. *Biochim. Biophys. Acta* **1757**, 886-912 (2006).
10. A. Siria *et al.*, Giant osmotic energy conversion measured in a single transmembrane boron nitride nanotube. *Nature* **494**, 455-458 (2013).
11. R. H. Tunuguntla, F. I. Allen, K. Kim, A. Belliveau, A. Noy, Ultrafast proton transport in sub-1-nm diameter carbon nanotube porins. *Nat. Nanotechnol.* **11**, 639-644 (2016).
12. J. Abraham *et al.*, Tunable sieving of ions using graphene oxide membranes. *Nat. Nanotechnol.* **12**, 546-550 (2017).
13. L. Chen *et al.*, Ion sieving in graphene oxide membranes via cationic control of interlayer spacing. *Nature* **550**, 380-383 (2017).
14. J. Feng *et al.*, Observation of ionic Coulomb blockade in nanopores. *Nat. Mater.* **15**, 850-855 (2016).
15. A. K. Geim, I. V. Grigorieva, van der Waals heterostructures. *Nature* **499**, 419-425 (2013).
16. B. Radha *et al.*, Molecular transport through capillaries made with atomic-scale precision. *Nature* **538**, 222-225 (2016).
17. A. Esfandiar *et al.*, Size effect in ion transport through angstrom-scale slits. *Science* **358**, 511-513 (2017).
18. See *Supplementary Information.*
19. B. Tansel, Significance of thermodynamic and physical characteristics on permeation of ions during membrane separation: Hydrated radius, hydration free energy and viscous effects. *Sep. Purif. Technol.* **86**, 119-126 (2012).
20. G. Algara-Siller *et al.*, Square ice in graphene nanocapillaries. *Nature* **519**, 443-445 (2015).
21. Y. Zhu, F. Wang, J. Bai, X. C. Zeng, H. Wu, Compression limit of two-dimensional water constrained in graphene nanocapillaries. *ACS Nano* **9**, 12197-12204 (2015).
22. M. Neek-Amal, F. M. Peeters, I. V. Grigorieva, A. K. Geim, Commensurability effects in viscosity of nanoconfined water. *ACS Nano* **10**, 3685-3692 (2016).
23. O. Björneholm *et al.*, Water at interfaces. *Chem. Rev.* **116**, 7698-7726 (2016).
24. S. Hu *et al.*, Proton transport through one-atom-thick crystals. *Nature* **516**, 227-230 (2014).
25. D. Marx, M. E. Tuckerman, J. Hutter, M. Parrinello, The nature of the hydrated excess proton in water. *Nature* **397**, 601-604 (1999).
26. M. Chen *et al.*, Hydroxide diffuses slower than hydronium in water because its solvated structure inhibits correlated proton transfer. *Nat. Chem.* **10**, 413-419 (2018).
27. F. Helfferich, *Ion Exchange*. (McGraw-Hill, New York, 1962).
28. J. Chen, A. Zen, J. G. Brandenburg, D. Alfè, A. Michaelides, Evidence for stable square ice from quantum Monte Carlo. *Phys. Rev. B* **94**, 220102 (2016).
29. I. Strauss, H. Chan, P. Král, Ultralong polarization chains induced by ions solvated in confined water monolayers. *J. Am. Chem. Soc.* **136**, 1170-1173 (2014).
30. K. A. Mauritz, R. B. Moore, State of understanding of nafion. *Chem. Rev.* **104**, 4535-4586 (2004).

# Supplementary Information

#1. Device fabrication

To make our 2D channels, we refined the procedures reported in the earlier work [16,17]. The starting point of the process was to make a free-standing silicon nitride (SiN) membrane using a commercial Si wafer covered with a ~500 nm thick SiN layer. A rectangular hole of ~ 3 µm × 190 µm in size was drilled through the membrane using photolithography and dry etching. This structure served as a mechanical support for the following van der Waals assembly (Fig. 1A of the main text). Three layers – bottom, spacer and top – made from monocrystals (purchased from *HQ Graphene*) of hexagonal boron nitride (hBN) or graphite were consecutively placed on top of the membrane. First, the bottom crystal (20 to 50 nm thick) was transferred to cover the hole in the SiN membrane. For the spacer, monolayer graphene was patterned into an array of parallel ribbons (~130 nm wide and spaced by ~130 nm) and then transferred on top of the bottom crystal. The ribbons were aligned perpendicular to the long axis of the rectangular hole. After that, dry plasma etching was employed from the back side of the SiN membrane to cut through the bottom and spacer layers extending the hole into the bilayer assembly. Following this step, a relatively thick (150 to 200 nm) graphite or hBN crystal was transferred on top, covering the resulting opening (Fig. 1A of the main text). It was important to have this crystal sufficiently thick to avoid its sagging, which would block the empty space between the spacers. The sagging was one of the reasons why no molecular transport could previously be detected for monolayer channels [16]. After each transfer, we annealed the assembly at 400°C in an argon-hydrogen atmosphere for 3 hours, which removed polymer residues and other contamination introduced during fabrication procedures. The annealing steps were found critical to get contamination-free (non-blocked) channels.

Cleaved graphite and hBN monocrystals often have rough edges with a gradually decreasing thickness. The thinning edges could sag inside the channels blocking their entries from the top side (Fig. 1A). This problem was particularly serious for monolayer channels because of their atomic-scale height. To remove the possible blockage by edge sagging, a gold strip (3 nm Cr/100 nm Au) was deposited on top using photolithography and thermal evaporation. The strip was arranged to run parallel to the opening in the SiN membrane (that is, parallel to the bottom entries of the 2D channels; see Fig. 1A of the main text). Then, plasma etching was again employed to remove edge regions of the top crystal, which effectively freshly opened the channels' top entries. The gold strip above the 2D channels defined their length $L$ and, also, served as a mechanical clamp [S1]. The latter was important to avoid delamination of the van der Waals assembly, which without the clamp often happened during electrical measurements in aqueous solutions.

The separation between two atomically-flat crystals of hBN or graphite, which are placed on top of each other without a crystallographic alignment, should in principle be larger than their interlayer spacing $a$ ($a \approx 3.4$ Å for both hBN and graphite) [S2-S5]. Moreover, if hBN and graphene crystallographic axes are carefully (within 2°) aligned, graphene placed on top hBN can exhibit out-of-plane corrugations of ~1 Å [S2-S5]. We avoided such 'moiré superlattice' effects by intentionally misaligning our crystals during their assembly [S6]. At larger angles, the distance between the stacked crystals is not expected to be notably different from the interlayer spacing. The increase should be no more than a fraction of Å, depending on alignment details [17,S7]. For brevity, this correction to $h$ is ignored in the discussions of the main text where the 2D capillaries are referred to as having $h \approx a$. Nonetheless, the correction may be important to quantitative modelling of the experimental results in future.

#2. Raman detection of water inside 2D channels

To corroborate that water filled our one-atom-high capillaries, we employed Raman spectroscopy. The devices used in these experiments were slightly modified with respect to those reported in the main text because SiN gave rise to a large background signal that obscured a feeble Raman signal from 2D water inside the channels. To this end, we used Si wafers with rectangular openings of ~ 8 µm × 35 µm in size (Supplementary Fig. 1A). The wafers were then covered with a 300 nm thick layer of $SiO_2$. The oxide layer was to facilitate the van der Waals assembly described in section 1 by enhancing the optical contrast of thin hBN and graphite crystals [S8]. The bottom layer in the Raman experiments was graphite whereas the top layer was optically-transparent hBN so that the laser light could reach water inside the channels.

The resulting monolayer devices (Supplementary Fig. 1A) were mounted on top of a container filled with deionized (DI) water. The Raman spectra were acquired at room temperature using a confocal Raman spectrometer with a green 532 nm laser (*Renishaw*). Spectra were collected from several locations as indicated in Supplementary Fig. 1A. Those locations were 1) liquid water inside the container underneath the rectangular hole, 2) Si wafer covered by the bottom graphite, 3) 2D channels without access to water so that they remained empty and 4) 2D channels adjacent to the opening, which could be filled with water. The former three locations were used as references. As expected, locations 1 and 2 exhibited the standard spectra of bulk water [S9,S10] and of graphite, respectively. The Raman spectrum of water had a characteristic broad peak at around 3200-3400 cm$^{-1}$ whereas graphite exhibited the well-known sharp peaks referred to as 2D, 2G and D+G. The Raman spectra acquired from locations 3 and 4 (empty and filled channels, respectively) also showed the 2D, 2G and D+G peaks because of the presence of bottom graphite. However, there is a clear difference between the Raman spectra of the empty and filled channels (see Supplementary Fig. 1B). The latter exhibited an additional broad peak at 3200-3400 cm$^{-1}$, the same peak as bulk water, which unambiguously proves the presence of liquid water inside the monolayer channels.

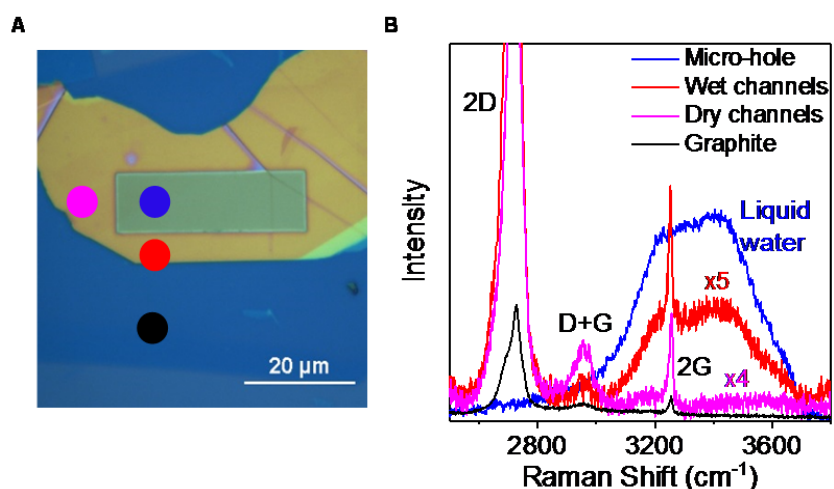

**Supplementary Fig. 1.** *Raman detection of two-dimensional water. (A) Optical image of the device used in our measurements. The opening in the Si/SiO$_2$ wafer, which is covered by an hBN crystal, shows up as a green rectangle (indicated by the blue spot). The rest of the top hBN layer is seen in dark yellow. The Si wafer appears in blue (bottom part of the image) whereas the graphite layer makes this blue somewhat darker (see the area around the black spot). The regions with empty and filled monolayer channels are marked by the pink and red spots, respectively. (B) Raman spectra from different locations in (A); color coded. The spectra acquired from the regions with the 2D channels are magnified as specified in the plot. Laser power of 1 mW; 100× objective lens; acquisition time of ~1 min. The laser beam had a diameter of ≥ 2 μm so that the Raman signal was collected from several channels.*

#3. Gravimetric measurements

To measure water permeation through our one-atom-high capillaries, we employed high-precision gravimetry with a resolution of about 1 μg and carried out recording over several days [16]. Importantly, even this precision would not be sufficient if we would not increase the water flux in our latest experiments (Fig. 1 of the main text) by increasing the number of channels to 1,500 and making them as short as possible ($L$ = 2 μm). The Si wafer with the capillary devices was mounted to seal a container filled with DI water (inset of Fig. 1C of the main text). The container was placed on a microbalance to monitor weight losses at a constant temperature (20 ± 0.1°C) and in a relative humidity of 30%. Both parameters were carefully controlled using an environmental chamber.

In our previous study [1], it was found that water permeation is a nonmonotonic function of the channel height, $h$. The maximum flow was observed for 4-layer channels (spacers made using $N$ = 4 layers of graphene). For the narrower slits with $N$ = 2 and 3, the water flow rate $Q$ decreased, which was attributed to an increase in the viscosity $\eta$ of confined water [16,22]. Our monolayer ($N$ = 1) channels reported in this work continued to exhibit the same trend such that $Q$ decreased further, by one order in magnitude with respect to bilayer ($N$ = 2) capillaries. This decrease is somewhat expected. Indeed, for long and wide channels with $w/h \gg 1$, the liquid flow rate driven by pressure $P$ is described by the Hagen-Poiseuille equation [16]

$$Q = \rho \frac{h^3}{12\eta}\left(1 + \frac{6\delta}{h}\right) P \frac{w}{L} \qquad (S1)$$

where $\rho$ is the water density and $\delta$ the slip length. The driving pressure $P$ is due to the evaporating external meniscus and was estimated to be of the order of 1000 bar for our angstrom-scale capillaries [16]. No significant changes are expected to occur in the meniscus if $h$ is below 1 nm and, in the first approximation, the pressure in eq. (S1) can be assumed constant. The same is also valid for the water density and the slip length [16,22]. The latter length was estimated to be ∼ 60 nm [16,S11]. This consideration implies that major changes in $Q$ could only be due to $h$ and $\eta$. Eq. (S1) yields the $h^2$ dependence, that is, 4 times decrease in $Q$ is expected for monolayer capillaries with respect to the bilayer ones. The additional discrepancy (by a factor of ∼ 2) can be accounted for by increased structural order in 2D water, which results in its increased viscosity $\eta$ [16,22,S12].

#4. Comparison of water flows through 2D channels and aquaporins

Unlike any artificial capillaries reported previously, our 2D channels exhibit selectivity similar to that of aquaporins and completely exclude ions. Therefore, it is instructive to compare the water permeation capacity of the two systems.

Aquaporins are protein channels in cell membranes with narrow constrictions from 2.8 Å to 10 Å in size. The narrowest ones match the effective size (∼2.8 Å) of individual water molecules and are also comparable to the height of our 2D channels, $h \approx 3.4$ Å. However, the 2D channels are much wider (up to 500 times; $w \approx 130$ nm) and longer (several microns as compared to 1 nm long aquaporins). Furthermore, the water flow through aquaporins is driven by typical pressures of ∼10 bar due to osmotic gradients across cell membranes [S13]. In the case of our 2D channels, the driving pressure $P$ is much larger, of the order of 1000 bar [16]. As a rough estimate, we use eq. (S1) and our experimental data (inset of Fig. 1C of the main text) to calculate the water flux if our artificial channels were as short as aquaporins but the flow was driven by much lower pressure gradients typical for aquaporins. As $Q \propto wP/L$ in eq. (S1), we find $Q \approx 6\times10^5$ g s$^{-1}$ m$^{-2}$ for our monolayer slits. This is two orders of magnitude larger than typical fluxes observed for aquaporins, $Q \approx 5\times10^3$ g s$^{-1}$ m$^{-2}$ [6,S13]. Despite obvious limitations (the two systems are quite different and even have different dimensionalities), the above comparison indicates that the reported 2D channels are efficient conduits of water, presumably due to a large slip length $\delta$ at atomically flat hydrophobic surfaces [16,S11].

#5. Ion transport measurements

In these experiments, the Si chip with 2D channels was placed to separate two Teflon reservoirs filled with various ionic solutions as shown in the inset of Fig. 2A of the main text. On both sides of the chip, we used polydimethylsiloxane seals to ensure that the 2D channels were the only conductive path between the reservoirs. This was crosschecked using blank SiN chips as well as the reference devices described in the main text, which had no channels. In both cases, we could detect only leakage currents $I$ of up to 5 pA at applied voltages $V$ of ∼ 0.1 V, presumably because of a finite conduction along the container walls. Two Ag/AgCl electrodes were used for electric measurements [17] and the results were regularly validated using standard reference electrodes (*HANA Instruments*). The applied voltages and resulting ionic currents were monitored using *Keithley* source meter (2636B). Typical *I-V* characteristics for different chloride solutions reported in the main text are shown in Supplementary Fig. 2A.

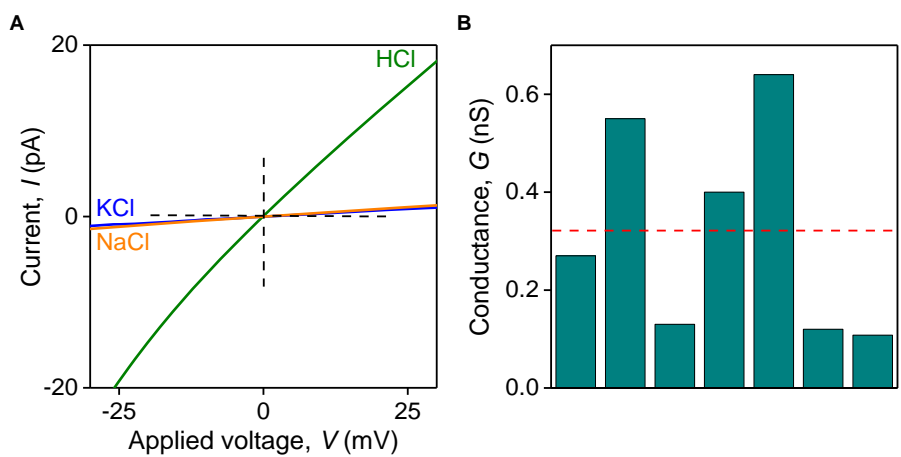

**Supplementary Fig. S2.** *Ion transport measurements for monolayer channels. (A) I-V characteristics for various 1M solutions using a monolayer device with n = 100 and L = 10 μm. Except for HCl, all other salts exhibited negligible conductance, undetectable within our resolution given by leakage currents of about 1 pA at 25 mV. Inset: Schematic of the experimental setup. (B) HCl conductance observed for seven devices such as in (A) using C = 1 M. The mean conductance is 0.32 ± 0.2 nS as indicated by the red line.*

Before each series of measurements for a given chloride solution, the reservoirs were sequentially washed with ethanol, ethanol:DI-water (50:50) and finally DI water, which ensured proper wetting of the 2D channels. For the concentration dependence measurements, we always started with DI water and then gradually increased the ion concentration, $C$. After finishing measurements for a particular chloride, we rinsed the cells with DI water several times, until the setup exhibited only the leakage conductance. Despite using the Au clamps described in section 1, our devices sometimes delaminated after electrical measurements at high $C$ and high $I$. Microscopy analysis revealed that the entire van der Waals assembly was lifted from the SiN membrane. Such damaged devices exhibited high conductance even in DI water and were discarded. The problem was particularly severe for the case of HCl where only seven (~40%) of the tested devices survived the entire test sequence from low to high $C$. The results for these devices are summarized in Supplementary Fig. 2B.

#6. Diffusion constant for protons in 2D water
The ion and proton conductivity σ inside our channels was calculated as $\sigma = G \times L \times A^{-1}$ where $G$ is the measured conductance, $L$ is the channel length and $A = n \times h \times w$ is the cross sectional area of the devices. σ is related to the mobility and concentration of ions or protons inside the channels through the expression (*44*) $\sigma = FC(\mu^+ + \mu^-)$ where $\mu^+$ and $\mu^-$ are the mobilities of $H^+$ and $Cl^-$, respectively, $F$ is the Faraday constant and $C$ is the ionic concentrations inside the channels [S14]. For our particular case where chlorine ions are excluded and only protons can transport through the channels, the above formula reduces to $\sigma = FC_p\mu^+$, where $C_p$ is the proton concentration. This yields the proton diffusion constant $d_p = (kT/e)\mu^+ = \sigma(kT/eFC_p)$. Because protons experience little barrier for entry into the 2D slits (see the main text), it is reasonable to assume that their concentration inside the channels is close to that in the outside HCl solution. Using the experimental data in Fig. 2B of the main text (red symbols), we find $d_p = 3.8 \pm 0.7 \times 10^{-6}$ cm$^2$ s$^{-1}$ (red curve) where the uncertainty arises from the standard error in the linear fit. Furthermore, the diffusion constant can be estimated using the measurements on Nafion devices, too. Despite the large uncertainty in the proton concentration in the latter case (see the main text), our data agree well with the above estimate for $d_p$.

#7. Non-monotonic dependence of proton diffusivity on dimensionality
In 3D liquid water, the number of hydrogen bonds around each water molecule is larger than 3 but less than 4 [S15], which means that there are plenty of defects in the hydrogen bond structure. These defects are crucial for rapid proton transport in water, which is consensually described by the Grotthuss model [25,S16]. It suggests that an oxygen atom at the defect site can accept an 'excess' proton and then pass it to the next water molecule, which involves the molecules' reorientations. The latter is a relatively slow process. In monolayer water, few structural defects are expected, according to molecular dynamics simulations [20,S17-S19]. One of the possible crystallographic arrangements of 2D water is shown in Supplementary Fig. 3A. This is so called square ice that may exist at room temperature [20,S19]. In this case, each water molecule has 4 hydrogen bonds, which makes it difficult to transfer excess protons between molecules using the Grotthuss mechanism. Another possible crystal structure of monolayer water is a hexagonal ice [S20], and the same arguments apply in this case, too. Other hydrogen-ordered configurations in 2D cannot be ruled out and require further studies. Another case of low-dimensional water is a 1D molecular chain shown in Supplementary Fig. 3B, where each water molecule has only two hydrogen bonds. This is expected to result in a higher probability to accept and pass excess protons, as argued before [5,11,S21]. Accordingly, proton conductivity in 1D water is expected and was observed to be much higher than in 3D liquid water [S21]. Our case of 2D water can be expected to exhibit the lowest proton conductivity, in agreement with the reported experiments that show $d_p$ to decrease by more than an order of magnitude with respect to 3D water and two orders with respect to 1D water.

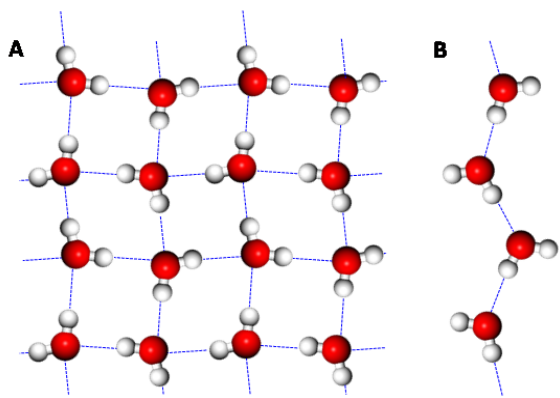

**Supplementary Fig. S3.** *Hydrogen bonding in 2D and 1D water. Schematic illustration of hydrogen bond networks in (**A**) monolayer water and (**B**) 1D single chain of water molecules [5,S21].*

**Supplementary references**

S1. A. Keerthi *et al.*, Ballistic molecular transport through two-dimensional channels. *Nature* **558**, 420-424 (2018).
S2. M. Neek-Amal *et al.*, Membrane amplitude and triaxial stress in twisted bilayer graphene deciphered using first-principles directed elasticity theory and scanning tunneling microscopy. *Phys. Rev. B* **90**, 064101 (2014).
S3. K. Uchida, S. Furuya, J.-I. Iwata, A. Oshiyama, Atomic corrugation and electron localization due to Moire patterns in twisted bilayer graphenes. *Phys. Rev. B* **90**, 155451 (2014).
S4. C. R. Woods *et al.*, Commensurate–incommensurate transition in graphene on hexagonal boron nitride. *Nat. Phys.* **10**, 451 (2014).
S5. H. Yoo *et al.*, Atomic reconstruction at van der Waals interface in twisted bilayer graphene. *arXiv:1804.03806* (2018).
S6. R. Krishna Kumar *et al.*, High-temperature quantum oscillations caused by recurring Bloch states in graphene superlattices. *Science* **357**, 181 (2017).
S7. K. Xu, P. Cao, J. R. Heath, Graphene Visualizes the First Water Adlayers on Mica at Ambient Conditions. *Science* **329**, 1188 (2010).
S8. R. V. Gorbachev *et al.*, Hunting for monolayer boron nitride: Optical and raman signatures. *Small* **7**, 465-468 (2011).

S9.     M. Erko, G. H. Findenegg, N. Cade, A. G. Michette, O. Paris, Confinement-induced structural changes of water studied by Raman scattering. *Phys. Rev. B* **84**, 104205 (2011).
S10.    M. Pastorczak, M. Kozanecki, J. Ulanski, Raman resonance effect in liquid water. *J. Phys. Chem. A* **112**, 10705-10707 (2008).
S11.    S. K. Kannam, B. D. Todd, J. S. Hansen, P. J. Daivis, How fast does water flow in carbon nanotubes? *J. Chem. Phys.* **138**, 094701 (2013).
S12.    O. Björneholm *et al.*, Water at Interfaces. *Chem. Rev.* **116**, 7698-7726 (2016).
S13.    B. Yang, A. S. Verkman, Water and glycerol permeabilities of aquaporins 1–5 and mip determined quantitatively by expression of epitope-tagged constructs inxenopus oocytes. *J. Biol. Chem.* **272**, 16140-16146 (1997).
S14.    A. J. Bard, L. R. Faulkner, *Electrochemical methods: Fundamentals and applications*. (John Wiley & Sons, Inc., New York, 2000).
S15.    R. Kumar, J. R. Schmidt, J. L. Skinner, Hydrogen bonding definitions and dynamics in liquid water. *J. Chem. Phys.* **126**, 204107 (2007).
S16.    M. Dominik, Proton transfer 200 years after von grotthuss: Insights from ab initio simulations. *ChemPhysChem* **7**, 1848-1870 (2006).
S17.    I. Strauss, H. Chan, P. Král, Ultralong polarization chains induced by ions solvated in confined water monolayers. *J. Am. Chem. Soc.* **136**, 1170-1173 (2014).
S18.    S. Jiao, Z. Xu, Non-Continuum Intercalated Water Diffusion Explains Fast Permeation through Graphene Oxide Membranes. *ACS Nano* **11**, 11152-11161 (2017).
S19.    Y. Zhu, F. Wang, H. Wu, Structural and dynamic characteristics in monolayer square ice. *J. Chem. Phys.* **147**, 044706 (2017).
S20.    J. Chen, G. Schusteritsch, C. J. Pickard, C. G. Salzmann, A. Michaelides, Two dimensional ice from first principles: structures and phase transitions. *Phys. Rev. Lett.* **116**, 025501 (2016).
S21.    C. Dellago, M. M. Naor, G. Hummer, Proton transport through water-filled carbon nanotubes. *Phys. Rev. Lett.* **90**, 105902 (2003).